\documentclass[aps,prd,amsmath,twocolumn,longbibliography]{revtex4-1}

\usepackage[dvips]{graphicx}
\usepackage{latexsym,epsfig,bm,psfrag,subfigure}
\usepackage{color}         
\usepackage[bookmarksnumbered,bookmarksopen,colorlinks,citecolor=blue,linkcolor=blue]{hyperref} %

\newcommand{\eq}[1]{Eq.~(\ref{#1})}

\def\bea{\begin{eqnarray}}
\def\eea{\end{eqnarray}}
\def\g{\gamma}\def\m{\mu}

\def\D{\Delta}
\def\e{\epsilon}\newcommand{\nn}{\nonumber\\&&}\def\L{\Lambda}
\begin{document}


\title{NT@UW-21-10\\Pions in Proton Structure and Everywhere Else }


\author{Mary Alberg$^{1,2}$}
\email{alberg@seattleu.edu}
\author{Lucas Ehinger$^1$}
\email{ehingerlucas@seattleu.edu}

\author{Gerald A. Miller$^2$}
\email{miller@uw.edu}
\affiliation{$^1$Department of Physics, Seattle University, Seattle, WA 98122, USA} 
\affiliation{$^2$Department of Physics, University of Washington, Seattle, WA  98195, USA}

\date{\today}

\begin{abstract}
The pion cloud is important in nuclear physics and in a variety of low-energy hadronic phenomena. We argue that  it is natural to expect it to also be important in 
 lepton-proton deep inelastic scattering and Drell-Yan studies of proton structure. We compute the necessary consequences of the pion cloud   in connection with the recent SeaQuest data. The effects are  detailed by using  the exact kinematics of the experiment. Good agreement with the measurements is obtained.  Thus the universality of pionic effects is understood.
   \end{abstract}




\maketitle
The recent striking  experimental finding~\cite{Dove:2021ejl} that anti-down quarks are more abundant in the proton than anti-up quarks  for all observed values of Bjorken $x$-variable   demands an interpretation and assessment of the consequences. This paper is aimed at providing such. 

The results of~\cite{Dove:2021ejl}  provide definitive experimental measurements of  the ratio $\bar d/\bar u$.
Although our early prediction~\cite{Alberg:2017ijg} using a pion cloud model is in qualitative agreement with that experiment, it is necessary  
to update the calculation by providing results for the specific kinematics of the experiment that are known only since the publication~\cite{Dove:2021ejl}.

We begin by explaining why  it is natural to expect that the pion cloud would play a role in probes of proton structure.
Pion exchange between nucleons    provides in the one pion exchange potential (OPEP)  the longest ranged component of the    strong force. It is an element of all models, from the ancient to the newest,  of the 
nucleon-nucleon interaction. The OPEP   is crucially responsible for the binding of nuclei~\cite{
Kaiser:1997mw,Kaiser:2001jx}.
Moreover, the presence of the pion as a significant component of the nuclear wave function is reinforced by the dominance of the pion in meson exchange corrections to a variety of nuclear properties. This was discussed long ago~\cite{Riska:1970jxh,Riska:1972zz} and recently~\cite{King:2020wmp}.

 If a nucleon emits a virtual  pion that is absorbed on another nucleon,  as in the OPEP, it can emit a pion that is absorbed by itself. This is because nucleons are identical particles and a pion can be absorbed on any nucleon. Thus, the nucleon must consist, at least part of the time, of a nucleon and a virtual pion. The very significant contributions of pions to nucleon and baryon properties have been well documented for a long time ~\cite{Theberge:1980ye,Thomas:1981vc,Theberge:1981mq,Theberge:1982xs,Bernard:1995dp}. Particular examples in which the pion cloud effects  are prominent are  the neutron charge distribution~\cite{Thomas:1981vc}, and baryon magnetic moments~\cite{Theberge:1982xs}.

Given that the proton wave function has $n\pi^+(u\bar d)$ and $p\pi^0$ components, with a two to one ratio of probabilities, there should be more anti-down quarks than anti-up quarks in the nucleon. This means that
the 
textbook description  that nucleons are composed of $u$ and $d$ valence constituent quarks,  cannot be the whole story.
Furthermore,
 the gluons inherent in QCD  generate quark-antiquark pairs via perturbative interactions. Thus one is led to the question: Do the pairs arise only from perturbative evolution at high momentum scales, or do they have a non-perturbative origin as in the pion cloud? A definitive answer would provide great help in understanding the nature of confinement and also fundamental aspects of the nucleon-nucleon force.  Perturbative QCD predicts a sea that is almost symmetric in light flavor. However, the discovery of the violation of the Gottfried sum rule told us that  $\bar{d}$ quarks are favored over  $\bar{u}$ quarks
\cite{Amaudruz:1991at} . This highlighted the importance of the pion cloud of the nucleon~\cite{Thomas:1983fh,Henley:1990kw}. Reviews are presented in  \cite{Speth:1996pz,Garvey:2001yq,Chang:2014jba,Geesaman:2018ixo}. More recent calculations of the difference $\bar d - \bar u$, the isovector component of the proton sea, have been published in \cite{Kofler:2017uzq,Cocuzza:2021cbi,McKenney:2015xis,Barry:2018ort}. We focus on the ratio $\bar d/\bar u$, determined by the SeaQuest experiment. The ratio has been a greater challenge for theory, since it depends on both the isoscalar and isovector components of the sea.

The concept of a component of a nucleon wave function makes sense only within a light front description of the nucleon.
Our previous formalism~\cite{Alberg:2017ijg} 
   provided a light-cone perturbation theory approach  capable of making   predictions with known uncertainties.
Previous  calculations had noted  ambiguities related to the dependence of the pion-baryon vertex function on momentum transfer and on the possible dependence upon the square of the four-momentum  of intermediate baryons, and much discussion ensued~\cite{Holtmann:1996be,Koepf:1995yh,Speth:1996pz,Strikman:2009bd,Strikman:2010pu,Alberg:2012wr,Ji:2009jc,Burkardt:2012hk,Ji:2013bca,Salamu:2014pka,Granados:2015rra,Granados:2016jjl}.   
Another more fundamental issue involving the loss of relativistic invariance occurs when the vertex function is treated   as depending on only three of the four necessary momentum variables. Our formalism  resolved both of these problems by using a four-dimensional formalism and by using experimental constraints on the pion-baryon vertex function.

 In a light-front formalism the proton wave function can be expressed as a sum of Fock-state components
 \cite{Lepage:1980fj,Brodsky:1997de,Brodsky:2000ii,Kovchegov:2012mbw}.     Our   hypothesis is that  the non-perturbative light-flavor sea  originates from  
  the bare nucleon,   pion-nucleon ($\pi N$) and  pion-Delta ($\pi\D$) components.
   The interactions are described by  using the relativistic  leading-order chiral Lagrangian \cite{Becher:1999he,Pascalutsa:1998pw}. 
  Displaying 
  the interaction terms to the relevant order in powers of the pion field, we use
 \bea
\mathcal{L}_{\rm int}& = 
 - \; {g_A\over 2 f_\pi} \bar\psi \gamma_\mu \gamma_5 \tau^a \psi 
\, \partial_\mu \pi^a 
 - {1\over f_\pi^2} \bar\psi \gamma_\mu \tau^a \psi \;
\e^{abc} \pi^b \partial_\mu \pi^c \nonumber\\&-{g_{\pi N\Delta}\over 2M} (\bar{\Delta}^i_\mu g^{\mu\nu}\psi\partial_\nu \pi^i +{\rm H. C.}) 
\label{lag}
\eea
 where $\psi$ is the Dirac field of the nucleon,   $\pi^a (a = 1, 2, 3)$
the chiral pion field and $M$ is the nucleon mass. In Eq.~(\ref{lag})
$g_A$ denotes the nucleon axial vector coupling and $f_\pi$ the 
pion decay constant.  The second  term  is the Weinberg-Tomazowa term which describes  
low-energy $\pi-$nucleon scattering. In the third term $g_{\pi N\Delta}$ is the $\pi N\Delta$ coupling constant, and the $\Delta^i_\mu$ field   is a vector   in both spin and  isospin space. 

 \begin{figure} [h]
\includegraphics[width=8.1991cm]{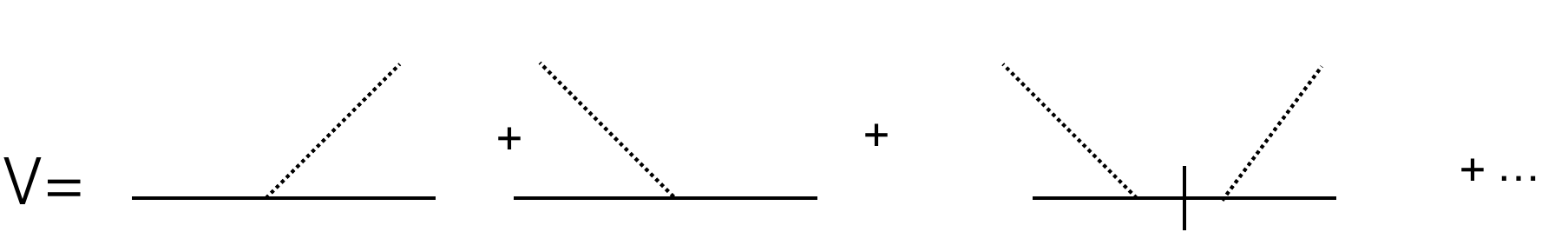}
\caption{Terms in the light-front Hamiltonian. }\label{Vpic}\end{figure}

    The light-front Hamiltonian operator is constructed   
   from the $T^{+-}$ component of the energy momentum tensor~\cite{Lepage:1980fj,Brodsky:1997de,Yan:1973qf,Miller:1997cr,Kovchegov:2012mbw}. The Hamiltonian can be written in terms of a sum of kinetic energy operators, $M_0^2$ and interaction  terms, denoted as $V$, see Fig.~\ref{Vpic}.  The first two terms are standard interactions, and the third is an instantaneous term  that     enters   only at 
   higher orders in the coupling constant.  The Hamiltonian forms of the  single-pion emission or absorption terms  (Fig.~\ref{Vpic}) are expressed as matrix elements evaluated between on-shell free nucleon spinors.  The light-front Schroedinger equation for the proton, $p$, is given by:
$ (M_0^2 +V)|p\rangle=M_p^2|p\rangle.$ 
To the desired second-order :
\bea |p\rangle\approx \sqrt Z |p\rangle_0+{1\over M_p^2-M_0^2}V|p\rangle_0,\label{wf}\eea
where $|p\rangle_0$ represents the  nucleon in the absence of the pion cloud, the bare nucleon, and $Z$ is a normalization constant. 
Given \eq{wf}, the wave function can be  expressed as a sum of Fock space components given  by 
\bea
|p\rangle=\sqrt Z |p\rangle_0+ \sum_{B=N,\D}\int d\Omega_{\pi B}|\pi B\rangle \langle \pi B|p\rangle_0, \label{fsw}
\eea
where $\int d\Omega_{\pi B}$ is a phase-space integral~\cite{Brodsky:2000ii,Kovchegov:2012mbw}.
In this formalism the  pion momentum distributions $f_{\pi B}(y)$, which represent the probability that a nucleon will fluctuate into a pion of light front momentum fraction $y$ and a baryon of light front momentum fraction $1-y$, are   squares of wave functions, $\left|\langle \pi B|\Psi\rangle\right|^2$ integrated over $k_\perp.$  \\

The Lagrangian of  \eq{lag}  is incomplete because it is not renormalizable. We  tame  divergences using a physically motivated set of regulators,  depending on 4-momenta, that are constrained by data.  
 If chiral symmetry  is maintained, 
 one finds that the $\pi N$ vertex function $ g_{\pi N}(t)$  and the nucleon axial form factor are related  by the 
generalized Goldberger-Treiman relation~\cite{Thomas:2001kw} (obtained with $m_\pi = 0$):
 \bea &
 Mg_A(t)=f_\pi g_{\pi N}(t),\label{relate}
\\&g_A(t)=g_A(0)/(1+(t/M_A^2))^2\label{MA} \eea where $t$ is the square of the four-momentum transferred to the 
 nucleon.  \eq{relate} follows from  partial conservation of the axial-vector current  (PCAC)  and the pion pole dominance of the pseudoscalar current. It is  obtained from  a matrix element of the axial vector current between two on-mass-shell nucleons.  
 The $t$-dependence of $g_A$ is determined for $t>0$ by low-momentum transfer experiments~\cite{Thomas:2001kw}, with $M_A$  the single parameter.   
   \eq{relate}   relates an essentially unmeasurable quantity $g_{\pi N}(t)$ with one $g_A(t)$ that is constrained by experiments.  The major uncertainty in previous calculations is largely removed. Some models, see e.g. \cite{Guichon:1982zk}, find differences between the $t-$dependence of $g_A(t)$ and $g_{\pi N}(t)$, which is allowed because $m_\pi \ne 0$.
   Uncertainties in the parameter $M_A$ are discussed in~\cite{Alberg:2017ijg}, where it is also shown that very large values of $t$  are not important in the calculations  of this paper.

In evaluating the nucleon wave function \eq{fsw} the necessary vertex function must be applicable to situations when either pion or baryon or both are off their mass shells.
 We use      frame-independent   pion-baryon form factors, in which a nucleon of mass $M$ and momentum  $p$ emits a pion of mass $\m$ and momentum $k$ and becomes a baryon of mass $M_B$ and momentum $p-k$:
 \bea &F(k , p, y)={\L^2 \over  k^2-\mu^2 -\L^2 +  i \e}\, {\L^2 \over {y\over 1-y}((p-k)^2-M_B^2)-\L^2+  i \e},\label{BigF}\eea
 where $y=k^+/p^+.$
   
 Using   $F(k,p,y)$ allows us to obtain a  pion-baryon  light front wave function.  The pion-nucleon component is given by:
\bea
& \Psi_{\rm a,LF}(k,p,s) 
 ={M g_A\over 2f_\pi (2\pi)^{3/2}}\sqrt{{y}\over  1-y} { \bar{u}(p-k)i\g^5\tau_a u(p)\over t+\mu^2} F_A(t),\nonumber\\&F_A(t)\equiv{2 \L^4\over \left(\L^2+t+\mu ^2\right) \left(2\Lambda ^2+t+\mu ^2\right)},\label{newr}
\eea
with $s$ and $a$ the spin and isospin labels for the proton.
 Expanding $F_A(t)$ to first order in $t$, comparing the result to the same expansion of  $g_A(t)/g_A(0)$ and matching the results determines the value  $\L=\sqrt{3}/2M_A$. Using  this $F(k,p,y)$  is equivalent to using a form factor of the form of \eq{relate}  in computing $f_{\pi N}(y)$~\cite{Alberg:2017ijg}.  The parameter-independence of this approach is maintained.\\

The pion 2D momentum distribution function $f_{\pi N}(y,t)$ is obtained by squaring  $\left|\Psi_{\rm a,LF}(k,p,s)\right|$ and summing over $a,s$. The result is
\bea  { f_{\pi N}(y,t)={3M^2\over 16\pi^2} {g_A^2\over f_\pi^2} y  {t\over(t+\m^2)^2} F_A^2(t)},
\label{nlf1} 
\eea
with  $t = (M^2y^2+k_\perp^2)/(1-y)$. 
The pion longitudinal momentum distribution function $f_{\pi N}(y)$ is then                    
\bea
f_{\pi N}(y) = \int_{t_N}^{\infty}{dt }f_{\pi N}(y,t)
\eea
 where $t_N\equiv{M^2y^2/(1-y)}$.
    Using  \eq{relate} or \eq{newr}  yields the same  $f_{\pi N}$ 
  because the integrand is dominated by the region of low values of $t$. 
  The pionic effects were shown to be of long-range~\cite{Alberg:2017ijg}
 by  studying the resulting  three-dimensional light-front  structure of the pion-baryon wave function.\\  
    
The intermediate $\D$ contribution is important because it is sizable and tends to favor $\bar u$ over $\bar d$.  We found~\cite{Alberg:2017ijg} that
 \bea&& 
  f_{\pi\Delta}(y)={1\over 12 \pi^2} \left({ g_{\pi N\Delta}\over 2M}\right)^2  y
\int_{t_\D}^\infty  dt\; {\;F_A^2(t)\over (t+\mu^2)^2}\nn\times
\left (t+{1\over4M_\Delta^2}(M^2-M_\Delta^2+t)^2\right)
((M+M_\Delta)^2+t) \nonumber\\&&\label{piD}
\eea
with $t_\D=(y^2M^2+y(M_\Delta^2-M^2))/(1-y)$.  \\

\begin{figure}[h]
\includegraphics[width=4.291cm]{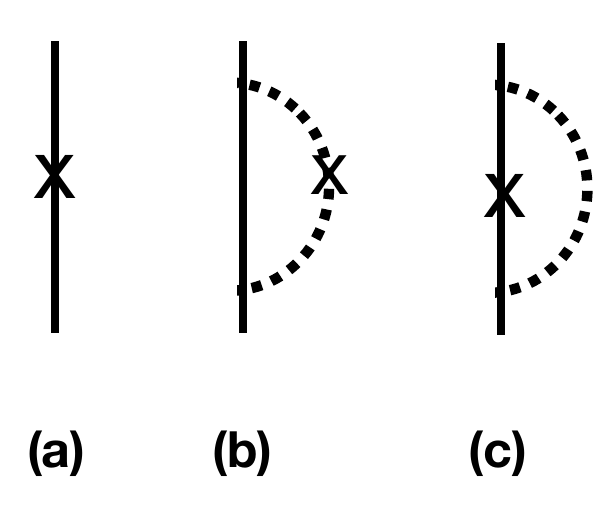}
\caption{(a) External interaction, X, with bare nucleon (solid line) (b)  External interaction, X, with the pion (c) External interaction, X, with the intermediate baryon. Here X represents the deep-inelastic scattering operator. }\label{fig:diag}\end{figure}

Next we use the Fock-space wave function of 
 \eq{fsw} to compute the  light flavor sea  component of the  
   nucleon wave function. Consider the role of the pion cloud in 
deep inelastic scattering,  (Fig.~\ref{fig:diag}).    One needs to include   terms in which the virtual photon hits (a) the bare nucleon, (b) the intermediate pion~\cite{Sullivan:1971kd}   and (c)  the intermediate baryon   $B$  of the $(\pi B)$ Fock-state component. 
 The key assumption of the present model is that   quantum interference effects involving different Fock space components are negligible
  because different final states obtained from deep  inelastic scattering  by the pion and by the nucleon are expected to be orthogonal.

The effects of the Weinberg-Tomazowa (WT)  term  vanish    because the  deep-inelastic scattering (DIS) operator, represented by  X in the figure is diagonal in the pion flavor index~\cite{Alberg:2017ijg}.  
 
Given the lack of interference effects, 
  one can represent at any $Q^2$ the  quark distribution functions of flavor $f = (\bar{u},\bar{d})$ in the nucleon sea as
\bea &q^f_N(x,Q^2)=Z q_{N0}^f(x,Q^2)+\sum_B f_{\pi  B}\otimes q_\pi^f +\sum_{B} f_{B\pi}\otimes q_B^f\nonumber\\\label{convol}\eea
 in which $B=N,\D$ and $ f_{\pi B}\otimes q_\pi^f \equiv \int_x^1{dy\over y} f_{\pi B}(y)q_\pi^f({x\over y},Q^2).$ 
The first symbol in the subscript represents the struck hadron, and the phase space factor in \eq{fsw} ensures that $f_{\pi B}(y)=f_{B\pi}(1-y)$, so that momentum is conserved.
The quark distributions of the hadrons in the cloud are given by $q_\pi^f(x,Q^2)$ and $q_B^f(x,Q^2)$, and the bare nucleon distributions are given by $q_{N0}^f(x,Q^2)$. 
The model is defined by the Fock-state expansion \eq{fsw} using meson-baryon states. The functions $f_{\pi B}(y)$ give the probability  that the proton fluctuates into a pion-baryon component as a function of the pion momentum fraction $y$. This defines the non-perturbative proton wave function that depends on the pion-baryon relative momentum
and is necessarily  independent of the momentum of any probe. This wave function is to be used to compute observables measured in reactions in which the probe interacts with the $\it hadrons$ in the Fock-state expansion of the proton. Pionic components make their presence known  in a variety of processes such as the computation of charge densities, magnetic moments and the nucleon-nucleon interaction.

 Next comes the issue of computing the structure functions obtained in deep-inelastic scattering and in the Drell-Yan process. This involves the photon-quark interaction. At the high momentum transfers relevant for those processes, the photon interacts with the quarks, not with the hadrons. The quark structure functions $q(x)$ are given in terms of the number of quarks in the hadron wave function at $q(\xi)$ and a function  $C$ that incorporates the dynamics of the photon-quark interaction,  in the schematic formula $q(x,Q^2) =\int d\xi q(\xi)C(x/\xi,Q^2)$\cite{Sterman:1993hfp}.
The function $C$ can be regarded as an effective photon-quark cross section~\cite{Halzen:1984mc}.  Thus the content of the model is that the evolution of the proton is contained in the evolution of the quarks that exist within the component hadrons.

 Evolution of the quark pdfs decreases the momentum fraction of the valence quarks as the momentum fractions of the sea quarks and gluons increase, but the momentum sum of all partons is still one.

We assume that  pionic fluctuations are the only source of the flavor asymmetry of the  proton sea. 
This is because all  quark-gluon processes are flavor independent if the quarks have the same mass. We have dominant $u,\bar{u}$ and $d,\bar{d}$ quarks that are essentially massless.  The bare proton and the intermediate $\D$ and nucleon pdfs have no contribution from pionic fluctuations, so they are flavor symmetric and we set $q^f_{N0}=q^f_\D=q^f_B$. We have suppressed the $Q^2$ dependence of $q^f_, q_{N0},q_\pi^f , q_B^f$ to simplify the notation.

  
Contributions to the antiquark sea of the proton come from the valence  and sea distributions of the pion $q_\pi^v$   and $q_\pi^s$
and the sea distributions $q_B^s$ and $q_N^s$ of the intermediate baryons and the bare proton. 
 The use of these distributions to describe deep inelastic scattering from a bound pion  follows from the light-front Fock space expansion, \eq{fsw}, which involves  only  on-mass-shell constituents. 
  With 
$
f_{\pi^+ n}={2\over 3}f_{\pi N},\: f_{\pi^0 p}={1\over 3}f_{\pi N},\:
f_{\pi^- \D^{++}}={1\over 2}f_{\pi \D},\: f_{\pi^0 \D^{+}}={1\over 3}f_{\pi \D},\:
 f_{\pi^+ \D^{0}}={1\over 6}f_{\pi \D},
$
the antiquark distributions are
\bea
&\bar{d}(x)= ({5\over 6}f_{\pi N}+{1\over 3}f_{\pi \Delta})\otimes q_\pi^v+ \bar{q}_{sym}(x)  \label{bard}\\
&\bar{u}(x)= ({1\over 6}f_{\pi N}+{2\over 3}f_{\pi \Delta})\otimes q_\pi^v+ \bar{q}_{sym}(x) \label{baru}\eea
where $ \bar{q}_{sym}(x)\equiv  \sum_ {B} f_{\pi B}\otimes q_\pi^s+\sum_ {B} f_{B \pi}\otimes q_B^s+ Z q_N^s(x).$
 The ${\pi N}$ terms favor the $\bar{d}$, but the ${\pi \D}$ terms favor the $\bar{u}$.

For comparison with SeaQuest results,  Eqs (11-13) require pion and baryon pdfs, $q_\pi^f(x,Q^2)$, $q_B^f(x,Q^2)$, and $q_{N0}^f(x,Q^2)$, 
 at the scale of the SeaQuest experiment, $Q^2 = 25.5$ GeV$^2$. We use pdfs determined by different groups from fits to experimental data.  We evolve these pdfs to the SeaQuest scale before using them in Eqs (12) and (13). This approach has been used in meson cloud models for many decades. e.g.  ~\cite{Koepf:1995yh,Holtmann:1996be, Salamu:2014pka}. We describe our evolution procedures below.

Two pion parton distributions were used: those of Aicher, Sch\"{a}fer and Vogelsang (ASV) \cite{Aicher:2010cb}, and the more recent pion pdfs of the xFitter Collaboration \cite{Novikov:2020snp}, which are consistent with the pion pdfs of the JAM Collaboration \cite{Barry:2018ort}, which have challenged the high-$x$ behavior of the ASV valence pdfs.
We evolved the ASV pion valence pdfs at next-to-leading order (NLO) from their starting scale of $Q_0^2 = 0.40$ GeV$^2$ to $Q^2 = 25.5$ GeV$^2$, the scale relevant for  SeaQuest. Our fit to the evolved valence distribution is given by
$q_\pi^v(x) =1.38\, x^{-0.320} (1-x)^{3.02 } \left(7.40\,  x^2+1\right).$ 
The ASV analysis used the pion sea quark pdfs of Gluck, Reya and Schienbein  \cite{Gluck:1999xe} at their starting scale. After NLO evolution to $25.5$ GeV$^2$, 
$q_\pi^s(x) = 0.113\, x^{-1.19} (1-x)^{5.10} \left(1-2.31\,
   \sqrt{x}+4.08\, x\right).$
We used the xFitter pdf parametrization from the LHAPDF6 Library with ApfelWeb \cite{Bertone:2013vaa,Carrazza:2014gfa} to evolve their pdfs at NLO to the SeaQuest scale.

Holtmann et al.~\cite{Holtmann:1996be} explained that the bare proton sea cannot  be determined directly from experimental data, which includes contributions from the pion cloud. Their model for bare proton parton distributions   \cite{Szczurek:1997fr}  used a fit to DIS data that included corrections to remove contributions from the pion cloud. In our updated calculation we used the program QCDNUM \cite{Botje:2010ay} 
to evolve at NLO their bare proton pdfs (valence, gluon and sea) from the starting scale of $Q_0^2 = 4$ GeV$^2$ to the SeaQuest scale of  $Q^2= 25.5$ GeV$^2$. This is the only change to our previous calculation of the bare sea.
We used the resulting bare sea distribution for the contributions of the proton in Fig. 2(a) and the intermediate baryons of Fig. 2(c).

 Other input parameters must be described before presenting numerical results. The pion-nucleon splitting function $f_{\pi N}(y)$ depends on the   coupling constant $g_{\pi N}$ and the form factor cutoff $\L$. The lower limit for $g_{\pi N}$  is 12.8, taken from the Goldberger-Treiman relation $g_{\pi N} ={M\over f_\pi}g_A$, with $g_A=1.267\pm0.04,\;M=0.939\; {\rm GeV},f_\pi=92.6\;{\rm MeV}$.   The upper limit is $g_{\pi N} =13.2$, consistent with the scattering data analysis of Perez et al. \cite{Perez:2016aol} and the muon-based determination of $g_A$ by Hill et al. \cite{Hill:2017wgb}.   As noted above the cutoff parameter of \eq{newr},  $\L=\sqrt{3}/2M_A$, is obtained at very low $t$.  The two resulting splitting functions are identical for all values of $y$, demonstrating that only small values of $t$ are important in the present calculations. In initial calculations we used the value
  $M_A = 1.03 \pm 0.04$ 
 GeV~\cite{Thomas:2001kw}. This early review result  was  confirmed by many authors~\cite{Bernard:2001rs,Juszczak:2009qa,Katori:2016yel,Nakamura:2016cnn,Meyer:2016oeg}, all obtaining results within the  stated  uncertainty. We have increased the uncertainty in our cutoff $\Lambda$ to $\pm 10\%$ to allow for a difference between the cutoffs in the $\pi NN$ form factor and the axial form factor. Although one early estimate, based on the cloudy bag model, suggested that the difference might be $\pm 20\%$ \cite{Guichon:1982zk},  later work using dispersion relations found consistency between the axial form factor cutoff and a  $\pi NN$ {\it monopole} cutoff
   of $\Lambda =0.80$ GeV $\pm 10\%$  \cite{Bockmann:1999nu,Ericson:2000md}. A monopole value of $\Lambda= $ 0.8 GeV corresponds to a dipole value of 1.1 GeV. 
 
\begin{figure}
\includegraphics[width=8.0cm]{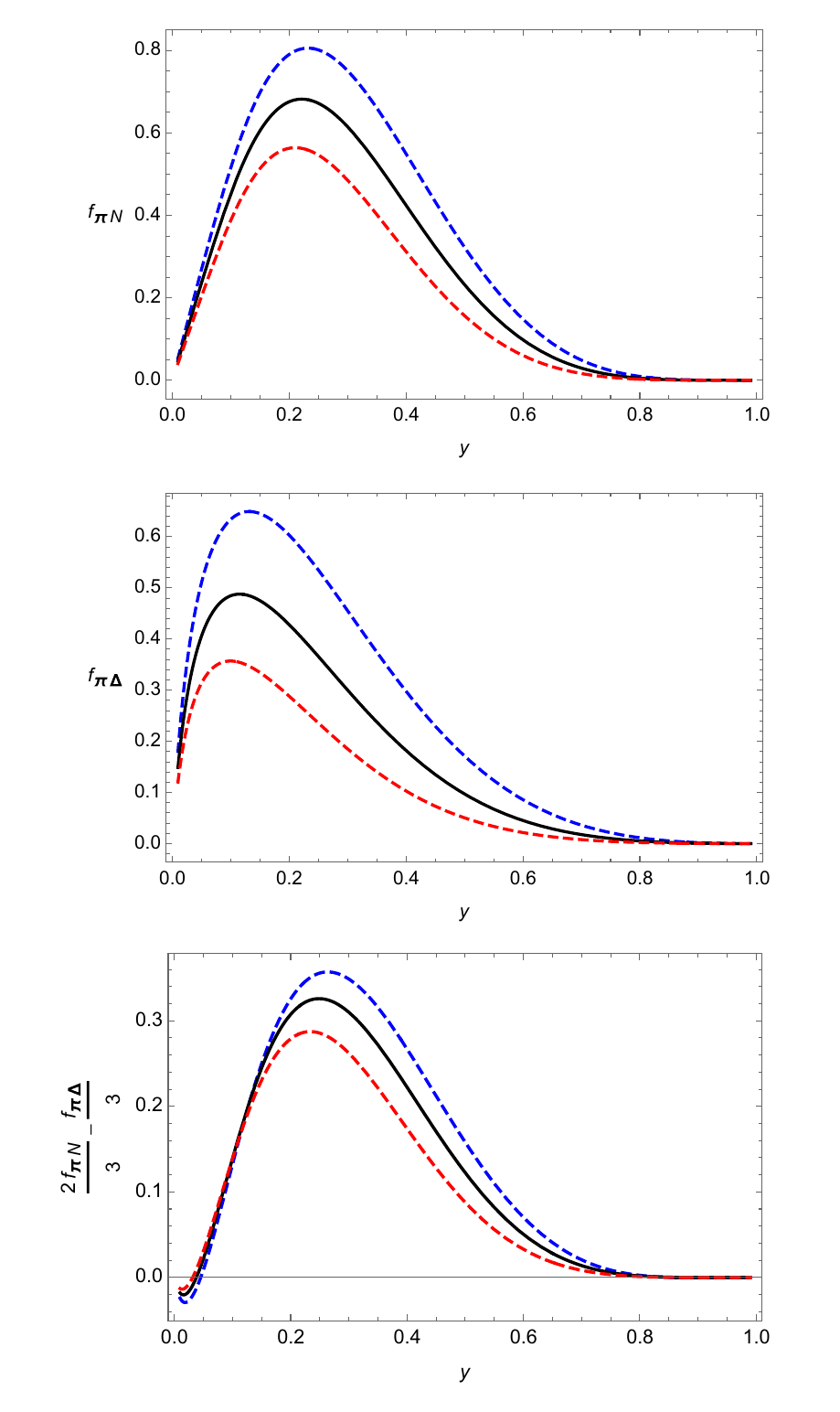}
\caption{Pion baryon splitting functions $f_{\pi {B}}(y)$, $B=N,\D$, are shown in the upper two panels. The solid lines are found using the central values of our coupling constants and cutoffs. The upper (blue) and lower (red) dashed lines are obtained using the maximum or minimum values, respectively,  of these parameters. The lowest panel shows the contribution of the splitting functions to the integrated asymmetry, $\bar{D}-\bar{U}$, \eq{intasym}. The smaller spread between the dashed lines is due to the correlation between the coupling constants and the use of the same cutoff in  $f_{\pi {N}}(y)$ and $f_{\pi {\D}}(y)$. }
\label{fig:splitfcns2}
\end{figure}

 
     The splitting function  $f_{\pi N}(y) $ 
      for a range of parameters bounded by the maximum and minimum values of $g_{\pi N}$ and $\L$ is shown in  Fig.~\ref{fig:splitfcns2}.

The value of the $\pi N\D$ coupling constant plays an important role in our calculations.  Both the upper limit $
({g_{\pi \D} \over g_{\pi N}})^2= {72\over 25}, 
\; g_{\pi \D}=1.7 g_{\pi N}$ and  lower limit,   obtained from the large $N_C$ limit of $g_{\pi \D}=1.5 \, g_{\pi N}$,
 are much smaller than the value $ g_{\pi \D}=2.2 g_{\pi N}$ extracted from the K-matrix analysis of~\cite{Oset:1981ih}. This difference is important because the contribution of the intermediate $\D$ is proportional to $g^2_{\pi \D}$.
The K-matrix analysis obtained $ g_{\pi \D}$ from the width of the $\Delta$ computed for the dominant $s$-channel diagram. This analysis is incomplete because it neglects the influence of the iteration of the crossed pion-nucleon Born term that makes a substantial contribution to the width. The importance of that diagram was explained in the textbook~\cite{Eisenberg:1980bf}. A detailed calculation of the pion-nucleon scattering in the (3,3) channel was made in~\cite{Theberge:1980ye}, which found a good description of the phase shift using quark-model values of the coupling constant.
That work used a static approximation, but Niskanen~\cite{Niskanen:1981pq} included higher-order effects which showed that the quark model values of the coupling constants can be consistent with the experimental width.  The definitive coupling constant compilation ~\cite{Dumbrajs:1983jd} found values consistent with the large $N_C$ limit. All of this early work was confirmed by recent calculations~\cite{Siemens:2016jwj,PhysRevD.87.054032,NavarroPerez:2014ovp} that find values of $g_{\pi \D}$ in accord (within errors) of the 
large $N_c$ calculations. Such values are consistent with the value extracted from the covariant $\D$ width at full one-loop
order~\cite{PhysRevD.87.054032} and with the extraction from $NN$ scattering~\cite{NavarroPerez:2014ovp}.  Moreover, our range of values of $g_{\pi \D}$ are used routinely in calculations of the nucleon-nucleon potential~\cite{Epelbaum:2008ga}. The net result is that the range of values of $g_{\pi \D}$ that we use is consistent with the width of the $\D$.

The splitting function $f_{\pi \D}(y)$ depends on the coupling constant $g_{\pi \D}$ and the form factor cutoff $\L$.   We use the same form factor and cutoff for $f_{\pi N}(y)$ and $f_{\pi \D}(y)$. 
The ratio $f_{\pi \D}(y)/f_{\pi N}(y) $ { is less than unity for the important regions of $y$. It does  increase as $y$ increases above 0.5, and becomes greater than unity at about $y=0.8$, where both splitting functions are vanishingly small.} Ref.~\cite{Alberg:2017ijg},  showed  in a detailed discussion that the splitting functions arise  from the long-range structure of the nucleon.

%
%

Finally,  it has been known for  a long time that the use of soft form factors (similar in range to those of the present study) leads to a convergent perturbation series~\cite{Theberge:1980ye,
Thomas:1981vc,Theberge:1981mq,AlvarezEstrada:1982bx} in the pion-baryon coupling constants..

Having justified the model, let's turn to the observations.
The integrated asymmetry  $\bar{D}-\bar{U}$ is the difference in number of $\bar d$ and $\bar u$ quarks in the proton sea. With $\bar{D} = \int_0^1 \bar{d}(x)dx, \,\bar{U} = \int_0^1 \bar{u}(x)dx$, the asymmetry is determined from \eq{bard} and \eq{baru} as 
\bea
\bar{D}- \bar{U} ={2\over 3}\int_0^1dy f_{\pi N}(y)-{1\over 3}\int_0^1dy f_{\pi \D}(y).
\label{intasym}
\eea
The experiment E866 \cite{Towell:2001nh} measured 
 $\bar{D}-\bar{U}=0.118 \pm 0.012$.  Our splitting functions predict 
 $0.098 \leq \bar{D}-\bar{U} \leq 0.131$, in excellent agreement with the experimental result.\\
\begin{figure}
\includegraphics[width=8.0cm]{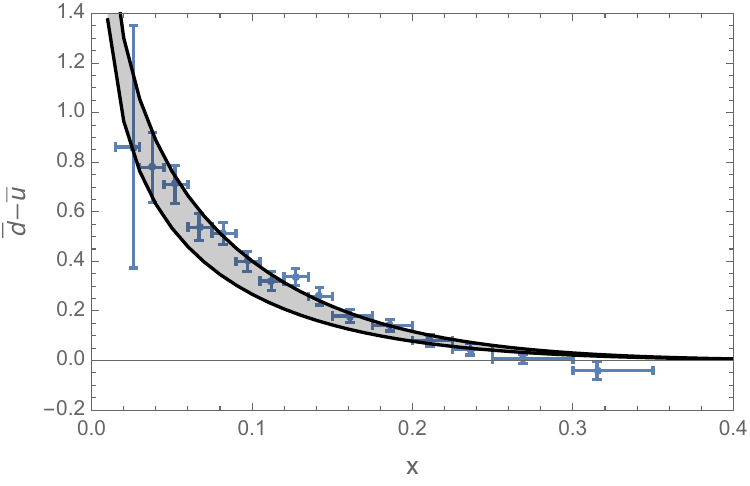}
\caption{$\bar{d}(x)- \bar{u}(x)$. Blue symbols from E866 \cite{Towell:2001nh}. The bands are computed using minimum and maximum values of the splitting functions shown in Fig.~\ref{fig:splitfcns2} in convolution with ASV or xFitter pion pdfs. }\label{fig:diff}\end{figure}
The computed values of  $\bar{d}(x)- \bar{u}(x)$ are compared to the E866 results 
 in  Fig.~\ref{fig:diff}, with bands obtained using minimum and maximum values of the splitting functions shown in Fig.~\ref{fig:splitfcns2}, in convolution with ASV or xFitter pion pdfs. The band includes the sum of the two contribution: $p \rightarrow \pi N$, and $p \rightarrow \pi \Delta$. The width of the band is narrow because of the correlation between the coupling constants: $g_{\pi \D} = r g_{\pi N}$, with $1.5 \leq r \leq 1.7$,  and the use of the same cutoff $\Lambda$ for both terms. This band is a definitive prediction of the present model.  We stress that in {\it any} model, $\bar u$ and $\bar d$ are correlated so that errors in each are partially cancelled in the ratio. 
  We find that a 15\% uncertainty  in $\bar{ d},\bar {u}$  at $x=0.3$ translates to 7\% in the ratio. 

\begin{figure}
\includegraphics[width=8.0cm]{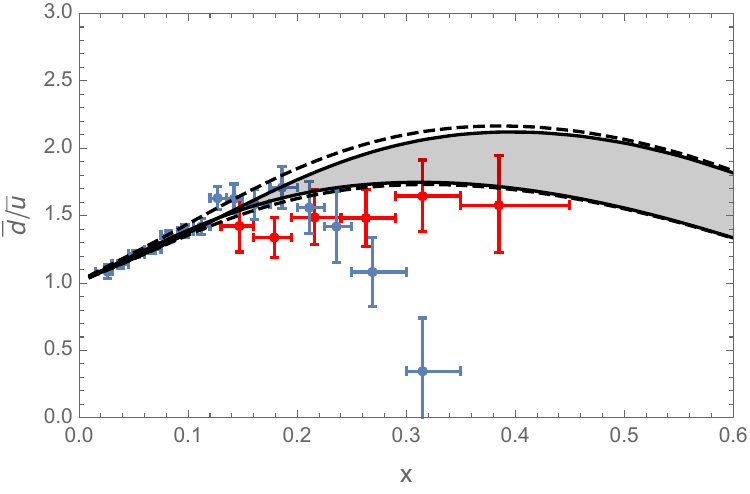}
\caption{$\bar{d}(x)/ \bar{u}(x)$ Blue symbols from E866 \cite{Towell:2001nh}. Red symbols are from SeaQuest \cite{Dove:2021ejl}.  The solid band is computed using minimum and maximum values of the splitting functions shown in Fig.~\ref{fig:splitfcns2} in convolution with ASV or xFitter pion pdfs, plus the bare sea of \cite{Szczurek:1997fr}. All pdfs were evolved to the SeaQuest scale of $Q^2 = 25.5$ GeV$^2$. The dashed lines include the effects of varying the bare sea by a factor of 0.75 or 1.25. 
}
\label{fig:ratio}

\end{figure}

Calculations of the   ratio $\bar{d}(x)/\bar{u}(x)$ are compared with   experimental data in  Fig.~\ref{fig:ratio}.  
 The results for values of $x$ less than about 0.15 arise from a combination of  pion cloud effects and the symmetric sea of the bare nucleon. 
  For larger values of $x$, terms of Fig.~2b dominate, with  the $\pi N$ contribution rising   with increasing $x$  until  $x \approx 0.34$. The ratio then drops  because of the enhancement of $\bar{u}$ (\eq{baru}) provided by   the   $\pi \Delta$ contribution, which becomes  relatively more  important as  $x$ increases.

Good agreement with experimental data is obtained  for $x < 0.2$, but the  decrease in the ratio $\bar{d}(x)/\bar{u}(x)$ found by E866 for higher values of $x$ is not reproduced. Our calculations are in agreement with the new results of SeaQuest, which show a slight rise in the ratio $\bar{d}(x)/\bar{u}(x)$ in the  $x$ domain covered by the experiment, with a slight hint of a downturn as $x \rightarrow 0.4$. Our calculation predicts that this would signal the increasing influence of the $\pi \Delta$ contribution. The prediction of Kofler and Pasquini for $\bar d/\bar u$ \cite{Kofler:2017uzq} lies well above ours because it does not include the $\pi \Delta$.
 
The present results for the ratio $\bar d/\bar u$ are a bit smaller than those of our earlier calculation~\cite{Alberg:2017ijg}, but have a wider uncertainty band because we considered two pion pdfs.. The lower ratio is caused by several factors, each of which increases $\bar d$ and $\bar u$. The dominant contribution to the proton sea comes from the valence antiquark distribution of the pion. The number of valence antiquarks is constant, but evolution increases $q^v_\pi (x)$  for $x < 0.2$ and decreases it for $x > 0.2$. In the latter domain, our evolution to the SeaQuest $Q^2=25.5$ GeV$^2$ yields higher $q^v_\pi (x)$ than our earlier evolution to the E866 $Q^2=54$ GeV$^2$, increasing the contributions of the first terms of Eqs. [\ref{bard},\ref{baru}]. The second terms of these equations make equal contributions to $\bar d$ and $\bar u$ from the symmetric sea quark pdfs of the pion and of the bare baryons. Evolution increases the total number of sea quarks. After convolution, the pion sea makes a small contribution to the proton sea, and its evolution has a small effect. In our earlier work we did not evolve the bare sea, so both $\bar d$ and $\bar u$ are increased in our present work. These increases in both numerator and denominator decrease the $\bar d/\bar u$ ratio, and bend our prediction band lower, improving its agreement with experiment. 

Our  2019 paper said ``The pion-baryon form factors  of our model are essentially model-independent, and the coupling constants are reasonably well-determined. For values of $x$ greater than about 0.15, the pion cloud effects dominate. The rise and then fall of the ratio $\bar{d}/\bar{u}$ are unalterable consequences of our approach. Significantly changing any of the input parameters would cause severe disagreements with other areas of nuclear physics, and would be tantamount to changing the model. If the high$-x$  E866 results  were to be confirmed by the SeaQuest experiment, the model would be ruled out."

It turned out that our predictions were in agreement with the SeaQuest data, even though we did not know the exact values of the kinematics. The present paper updates the earlier calculation by including evolution of the bare nucleon sea and using the now known SeaQuest kinematics. 
The present calculations show that the changes produce small effects, and further
that our earlier prediction is improved to a  good reproduction of the data.

The formalism presented here shows how to properly  obtain pion-baryon vertex functions  
 in a four-dimensional treatment that  includes the effects of the  uncertainties in the input parameters in a controlled fashion.  
 Our result is a chiral light-front perturbation theory calculation of the  wave function that successfully describes  the flavor content of the nucleonic  light-quark sea. 
 
 This shows that pionic effects are here, there and everywhere.
 
 \section*{Acknowledgements}
The work of M.A. and L.E. was supported by the Research in
Undergraduate Institutions Program of the US National Science
Foundation under Award No. 2012982.
The work of G.A.M. was supported by the
{ USDOE}  Office of Science, Office of Nuclear
Physics under Grant No. DE-FG02-97ER41014.

\bibliography{pion}

\end{document}